\documentstyle[preprint,aps,epsfig]{revtex}
\def\phirho{\phi \rightarrow \rho \gamma \gamma}
\def\phiom{\phi \rightarrow \omega  \gamma \gamma}
\begin{document}
\draft
\preprint{
SNUTP 95-083}
\bigskip
\title{ Chiral perturbation theory vs. vector meson dominance \\
in the decays $\phi \rightarrow \rho \gamma \gamma$
and $\phi \rightarrow \omega  \gamma \gamma$}
\author{ Pyungwon Ko$^{1,}$\footnote{ pko@phyb.snu.ac.kr}}
\address{$^{1}$Department of Physics, Hong-Ik University,
Seoul 121-791, Korea }
\author{Jungil Lee$^{2,}$\footnote{jungil@fire.snu.ac.kr}
    and H.S. Song$^{2,}$\footnote{hssong@phyy.snu.ac.kr} }
\address{
$^{2}$Department of physics and Center for Theoretical Physics
\\ Seoul National University,
Seoul 151-742, Korea  }
\date{\today}
\maketitle
\begin{abstract}
It is pointed out that
the radiative decays of a $\phi$ meson,
$\phirho$ and $\phiom$, receive dominant contributions from the
pseudoscalar ($P = \eta, \eta^{'}$) exchanges.
Using the vector meson dominance model, we find that
$B (\phirho) \approx 1.3 \times 10^{-4}$ and $B (\phiom) \approx 1.5
\times 10^{-5}$, which are mainly from the $\eta^{'}$ pole.
Thus, these decays are well within the reach of the $\phi$ factory.
Our estimates are a few orders of magnitude larger than the chiral
loop contributions
in the heavy vector meson chiral lagrangian, which is about ( a few )
$\times 10^{-9}$.
\end{abstract}

\narrowtext


1. Radiative decays of $\phi$ mesons, $\phirho$ and $\phiom$, are
rare processes, and thus have never been observed yet.
However, these may be observed
in the $\phi$ factory at Frascati where $10^{11}$ $\phi$'s would be produced
per year.
In this vein, it would be interesting to estimate the branching ratios for
these decays in a reasonable way.

In a different context, Leibovich {\it et al.} have recently
considered these decays in the heavy vector meson chiral
lagrangian approach  \cite{wise}.  In this framework, there are two
contributions to these processes : the loop (pseudoscalar-vector meson
loop) and the anomaly term from $\phi \rightarrow \rho (\omega) + \pi^{0}$
followed by $\pi^{0} \rightarrow \gamma \gamma$.  The loop contribution
depends on a parameter $g_2$ (the $VVP$ coupling) which enters in the
heavy vector meson chiral lagrangian \cite{wise} :
\begin{eqnarray}
B(\phirho)_{loop} & = & 5.8 \times 10^{-9}~\left( {g_{2} \over 0.75}
 \right)^4,
\\
B(\phiom)_{loop} & = & 4.2 \times 10^{-9}~\left( {g_{2} \over 0.75}
\right)^4,
\end{eqnarray}
where the results are normalized for $g_{2} = 0.75$ as predicted in the
chiral quark model.
For $\phiom$, the anomaly contribution from  $\phi \rightarrow \omega
\pi^0$ is negligible compared to the loop contribution shown above.
On the other hand,  the anomaly term
is not negligible for the other decay $\phirho$. However, one
can still find a region in the phase space where the loop
contributions dominate
the anomaly contributions. Since the loop contributions to $\phirho$ and
$\phiom$ are finite to the order considered in Ref.~\cite{wise},
the authors of Ref.~\cite{wise} suggested that one may relate the
measured branching ratios of $\phirho$ and $\phiom$ with Eqs.~(1) and (2)
in order to fix the low energy constant $g_2$ in the heavy meson
chiral lagrangian.

In this letter, we point out that 
there is an important class of Feynman diagrams
(see Fig.~1) which has been neglected in Ref.~\cite{wise}.
Feynman diagrams shown in Fig.~1 cannot occur in the heavy vector
meson chiral lagrangian, since the heavy vector
meson number is to be conserved in the chiral lagrangian
approach \cite{wise0}
(in the absence of weak interactions).   However, a vector meson can be
either created or destroyed
via electromagnetic interaction
({\it e.g.} $\omega \rightarrow
\pi^{0} + \gamma$, and $\eta^{'} \rightarrow \rho^{0} + \gamma$, etc.),
unlike the heavy quarks or the heavy baryons whose
numbers do not change by electromagnetic interactions.
Therefore, there is no apparent reason why Fig.~1 can be ignored in
$\phirho$ and  $\phiom$, and it is the purpose of this letter to calculate
the contributions of Feynman diagrams shown in Fig.~1
in the usual vector meson dominance model.
In fact, it  has long been known that the similar diagrams
are the most important in the decay $\eta \rightarrow
\pi^{0} \gamma \gamma$
through the $\rho$ and $\omega$ exchanges \cite{ko}.
\vspace{.3in}

2. The $VP\gamma$ vertex can be obtained from the following phenomenological
lagrangian (which can be also derived from the chiral lagrangian with
vector mesons in the hidden symmetry scheme \cite{fujiwara}) :
\begin{equation}
{\cal L} (V P \gamma) = {e\over g}~g_{\omega \rho \pi}~\epsilon_{\mu \nu
\alpha \beta}~F^{\mu\nu}~{\rm Tr}~\left[ P \left\{ Q, \partial^{\alpha}
V^{\beta} \right\} \right],
\end{equation}
where $g = g_{\rho\pi\pi} = 5.85$ using the KSFR relation, and
\begin{equation}
g_{\omega \rho \pi} = - {3 g^{2} \over 8 \pi^{2} f_{\pi}},
\end{equation}
with $f_{\pi} = 93$ MeV being the pion decay constant.
The matrix $Q = {\rm diag} (2/3, -1/3, -1/3)$ is the electric charges of three
light quarks, and $P$ and $V$ are the $3 \times 3$ matrix fields for
pseudoscalar and vector meson nonets :
\begin{eqnarray}
P&=&
\frac{1}{\sqrt{2}}
\left(
\begin{array}{ccc}
 \frac{1}{\sqrt{2}}\pi^0
 +\frac{1}{\sqrt{6}}\eta_8
 +\frac{1}{\sqrt{3}}\eta_0 & \pi^+ & K^+ \\
 \pi^- &
 -\frac{1}{\sqrt{2}}\pi^0
 +\frac{1}{\sqrt{6}}\eta_8
 +\frac{1}{\sqrt{3}}\eta_0 & K^0\\
 K^- & \bar{K^0} &
 -\frac{2}{\sqrt{6}}\eta_8+\frac{1}{\sqrt{3}}\eta_0
 \end{array}
 \right),\\
 V&=&
 \frac{1}{\sqrt{2}}
 \left(
 \begin{array}{ccc}
 \frac{1}{\sqrt{2}}(\rho^0+\omega)    &\rho^+   &K^{*+}\\
 \rho^-    &\frac{1}{\sqrt{2}}(-\rho^0+\omega)  &K^{*0}\\
 K^{*-}    &\bar{K}^{*0}                        &-\phi
 \end{array}
 \right).
 \end{eqnarray}
We assume the ideal mixing for $\omega$ and $\phi$, and a
partial mixing for $\eta$ and $\eta^{'}$ :
\begin{eqnarray}
\eta & = & \eta_{8} ~\cos \theta_{p} - \eta_{0} ~\sin \theta_{p},
\\
\eta^{'} & = & \eta_{8} ~\sin \theta_{p} + \eta_{0} ~\cos \theta_{p},
\end{eqnarray}
with $\theta_{p} \simeq - 20^{\circ}$.

One can extract the ~$V P \gamma
{}~(V=\phi,\rho,\omega$~and~$P=\pi^0,\eta,\eta^\prime$)
vertices from the above effective lagrangian, defining $C_{PV}$
as follows :
\begin{equation}
{\cal M}_{V P \gamma}=\frac{e}{g}g_{_{\omega \rho \pi}}
C_{PV}
\epsilon_{\mu\nu\alpha\beta} k^\alpha \epsilon^\beta (\gamma)
p^\mu \epsilon^\nu (V)
\end{equation}
where $k$ and $p$ are the momenta of $\gamma$ and $V$ respectively.
Explicit values of the coefficients $C_{PV}$'s are
given in Table ~I.

Let us first check how good this $SU(3)_V$ symmetric interaction lagrangian
is by considering various decays, $V \rightarrow P \gamma$ and $P \rightarrow
V \gamma$.
The decay rate for $i \rightarrow f + \gamma$ described by the above vertex
is given by
\begin{equation}
\Gamma ( i \rightarrow f + \gamma ) = {1\over (2 S_{i} + 1)}~{g^{2} \over
4 \pi}~C_{PV}^{2}~{9~ \alpha \over 128 \pi^3 f_{\pi}^2}~M_{i}^{3}~
\left[ 1 - {M_{f}^2 \over M_{i}^2} \right]^{3},
\end{equation}
where $S_i$ is the spin of the initial particle, $i$.

In Table~II, we present our predictions (using the $C_{PV}$'s in Table~1)
along with the measured  branching ratios.
The agreements between the two are reasonably
good, except for $\phi \rightarrow \pi^{0} \gamma$, which is OZI-forbidden
decay and thus is of higher order in $1/N_c$.  Also, note that our
prediction, $B(\phi \rightarrow \eta^{'} \gamma) = 2.1 \times 10^{-4}$
is a factor of two below the current upper limit ($< 4.1  \times 10^{-4}$).
Thus, this decay may be observed soon at the $\phi$ factory.  Since
$B(\eta^{'} \rightarrow \rho \gamma) = (30.2 \pm 1.3) \%$, this decay
($\phi \rightarrow \eta^{'} \gamma$) followed by $\eta^{'} \rightarrow
\rho \gamma$  can  constitute a large component of $\phirho$.
The results shown in Table~I suggest that the $VP\gamma$ interaction
lagrangian, Eq.~(3),  may
be used in studying other processes such as $\phirho$ and $\phiom$.
\vspace{.3in}

3. Using the $VP\gamma$ vertices in Table~I, it is straightforward to
calculate Feynman diagrams shown in Fig.~1, and get the decay rates and
the $\gamma \gamma$ spectra in $\phirho$ and $\phiom$.
It is convenient to use the following variables :
\begin{eqnarray}
s & \equiv & ( P_{\phi} - P_{\rho} )^{2} = ( k + k^{'} )^2,
\\
t & \equiv & ( P_{\phi} - k )^{2} = ( P_{\rho} + k^{'})^2,
\\
u & \equiv & ( P_{\phi} - k^{'} )^{2} = ( P_{\rho} + k )^{2},
\end{eqnarray}
with $s + t + u = M_{\phi}^{2} + M_{\rho}^2$.
The allowed ranges for $s$ and $t$ are
\begin{eqnarray}
0 \leq s \leq (M_{\phi} - M_{\rho})^{2}, ~~~~~~~~~~~~~~~~~
t_{0} \leq t \leq t_{1},
\\
t_{0,1} = {1\over 2}~\left[ \left( M_{\phi}^{2} + M_{\rho}^{2} - s \right)
 \mp  \sqrt{ ( M_{\phi}^{2} + M_{\rho}^{2} - s )^{2} - 4 M_{\rho}^{2}
M_{\phi}^{2} } ~\right].
\end{eqnarray}
In terms of $s,t,u$ variables, the amplitude for $\phirho$ is given by
\begin{eqnarray}
{\cal M}&=&-(\frac{e}{g}g_{_{\omega \rho \pi}})^2
\sum_{P=\eta,\eta^\prime}
c_{_{PV_1}} c_{_{PV_2}}
\left(
\frac{\epsilon_{\mu\alpha\kappa \sigma}
      \epsilon_{\nu\beta \lambda\tau  }
     }{t-m_{_P}^2+i~m_{_{P}}\Gamma_{P} }
+
\frac{\epsilon_{\mu\beta \lambda\sigma}
      \epsilon_{\nu\alpha\kappa \tau  }
     }{u-m_{_P}^2+i~m_{_{P}}\Gamma_{P} }
\right)\nonumber\\
&&\times~
k_1^\kappa~
k_2^\lambda~
p_1^\sigma~
p_2^\tau~
\varepsilon^{*\mu}(k_1)
\varepsilon^{*\nu}(k_2)
\varepsilon^{\alpha}(p_1)
\varepsilon^{*\beta}(p_2)      \\
&=&-
\left(
  f_t  \epsilon_{\mu\alpha\kappa \sigma}
       \epsilon_{\nu\beta \lambda\tau  }
+ f_u  \epsilon_{\mu\beta \lambda\sigma}
       \epsilon_{\nu\alpha\kappa \tau  }
\right)   
k_1^\kappa~
k_2^\lambda~
p_1^\sigma~
p_2^\tau~
\varepsilon^{*\mu}(k_1)
\varepsilon^{*\nu}(k_2)
\varepsilon^{\alpha}(p_1)
\varepsilon^{*\beta}(p_2)
\end{eqnarray}
where
\begin{eqnarray}
f_t&=&
(\frac{e}{g}g_{_{\omega \rho \pi}})^2
\sum_{P=\eta,\eta^\prime}
\frac{ c_{_{PV_1}} c_{_{PV_2}}  }
     {t-m_{_P}^2+i~m_{_{P}}\Gamma_{P} }
\nonumber\\
f_u&=&
(\frac{e}{g}g_{_{\omega \rho \pi}})^2
\sum_{P=\eta,\eta^\prime}
\frac{ c_{_{PV_1}} c_{_{PV_2}}  }
     {u-m_{_P}^2+i~m_{_{P}}\Gamma_{P} }
\end{eqnarray}
where $V_1=\phi$ and $V_2=\rho($  or~$\omega)$.
For $P = \eta^{'}$, the intermediate propagator can develop a pole for
$t$ (or $u$) $= m_{\eta^{'}}^2$, and we have to include its decay width in
the denominator of the propagator for $\eta^{'} $ :
$\Gamma_{\eta^\prime} = ( 0.198 \pm 0.019 ) \mbox{MeV}$.

The square of the above amplitude, when averaged over the initial spin,
and summed over the final spins, is simplified as follows :
\begin{eqnarray}
\overline{|{\cal M}|^2}
&=& \frac{1}{2}\cdot\frac{1}{3}
\sum_{\mbox{spin}}|{\cal M}|^2\nonumber \\
&=& \frac{1}{24}
\left[~
 |f_t|^2 (t-m_1^2)^2 (t-m_2^2)^2
+|f_u|^2 (u-m_1^2)^2 (u-m_2^2)^2\right.\nonumber\\
&& \left.  \hskip .5cm
{}~~+\mbox{Re}(f_t f^*_u)( s^2 m_1^2 m_2^2 +(m_1^2 m_2^2-t u)^2 )
{}~\right].
\end{eqnarray}
Here we have included the factor of $1/2$ in order to take into account
two identical particles (two photons) in the final state.
The decay rate can be obtained by integrating the following expression
over the variable $t$ :
\begin{equation}
{ d \Gamma \over d m_{\gamma \gamma}^{2}} =
{1\over (2 \pi)^3}~{1 \over 32 M_{\phi}^3}~\int_{t_0}^{t_1}~
\overline{ \left| {\cal M} \right|^{2} }~ dt.
\label{eq:dgds}
\end{equation}
After numerical integrations, we get
\begin{eqnarray}
B(\phirho) & = & 1.3 \times 10^{-4},
\\
B(\phiom)  & = & 1.5 \times 10^{-5}.
\end{eqnarray}
These are much larger compared to the loop contributions,
Eqs.~(1) and (2), obtained in the framework of the heavy vector meson
chiral lagrangian approach \cite{wise}.
Hence, the claim that these decays might be useful in constraining the
coupling $g_2$ in the heavy vector meson chiral lagrangian may not be
viable \footnote{
There are also  contributions from the anomaly via
$\phi \rightarrow \rho ({\rm or}~\omega) + \pi^{0} ({\rm or}~\eta, \eta^{'})
$ followed by $\pi^{0}, \eta, \eta^{'} \rightarrow \gamma \gamma$.  These
decays are dominated by the $\pi^0$ contribution, and can be suitably
removed by imposing a cut on $m_{\gamma \gamma} \approx  m_{\pi^0}$.
Hence, we do not consider these possibilities in this work.}.

One can also study the $m_{\gamma \gamma}$ spectra in $\phirho$ and $\phiom$
from Eq.~(\ref{eq:dgds}) as a function of $m_{\gamma \gamma}$, as shown
in Fig.~2.
The two spectra are the same except for (i) the overall normalization of
a factor $1/9$ from different Clebsch-Gordan
coefficients (as shown in Table~I), and (ii) the slight mass difference
between $\rho$ and $\omega$.  Both decays are dominated by $\phi \rightarrow
\eta^{'} \gamma$ followed by $\eta^{'} \rightarrow \rho ({\rm or}~\omega)
\gamma$.
\vspace{.3in}

4. In summary, we have considered the radiative decays of $\phi$ mesons,
$\phirho$ and $\phiom$ in the vector meson dominance model.  From
the usual $VP\gamma$ vertices, we find that these decays occur at the
level of $1.3 \times 10^{-4}$ and $1.5 \times 10^{-5}$ in the branching
ratios, respectively, and thus should be within
the reach of the $\phi$ factory at Frascati.
Also, these contributions
are dominant in size over the loop contributions considered in the
framework of the
heavy vector meson chiral perturbation theory \cite{wise},
and pose a doubt on the claim made in Ref.~\cite{wise}.
Finally, our model predicts that $B(\phi \rightarrow \eta^{'}
\gamma) = 2.1 \times 10^{-4}$,
which is just a factor of two below the current upper limit
and thus can be easily tested in the near future.
This decay actually dominates $\phirho$ and $\phiom$ in our model.
Thus, detection of this decay at the predicted level would constitute
another test of our model based on the vector meson dominance.
\vspace{.3in}

\acknowledgements

This work was supported in part by KOSEF through CTP at Seoul National
University. P.K. is supported in part by the Basic Science Research
Program, Ministry of Education, 1994, Project No. BSRI--94--2425.

\begin{figure}
\end{figure}
\vskip 1cm
\begin{figure}
\begin{center}
\epsfig{file=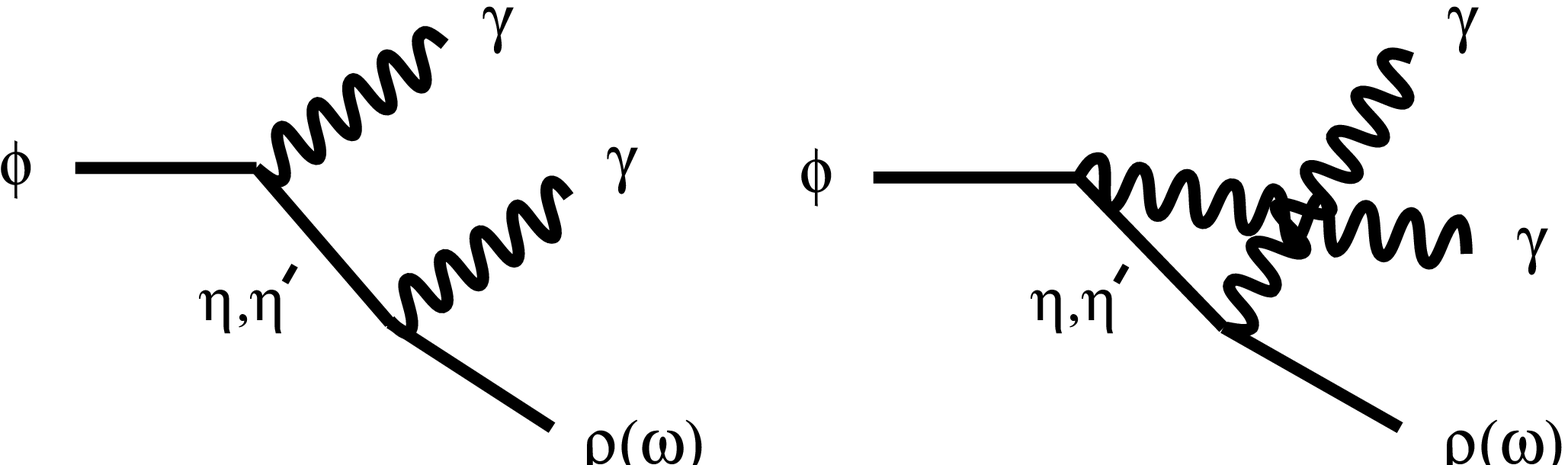,width=14cm,height=4cm}
\end{center}
\vskip 1cm
\caption{Feynman diagrams for $\phirho$ and $\phiom$ in the vector meson
dominance model.}
\label{figone}
\end{figure}
\vskip 5cm
\begin{figure}
\begin{picture}(200,100)(-50,0)
\put(0,0){
\epsfig{file=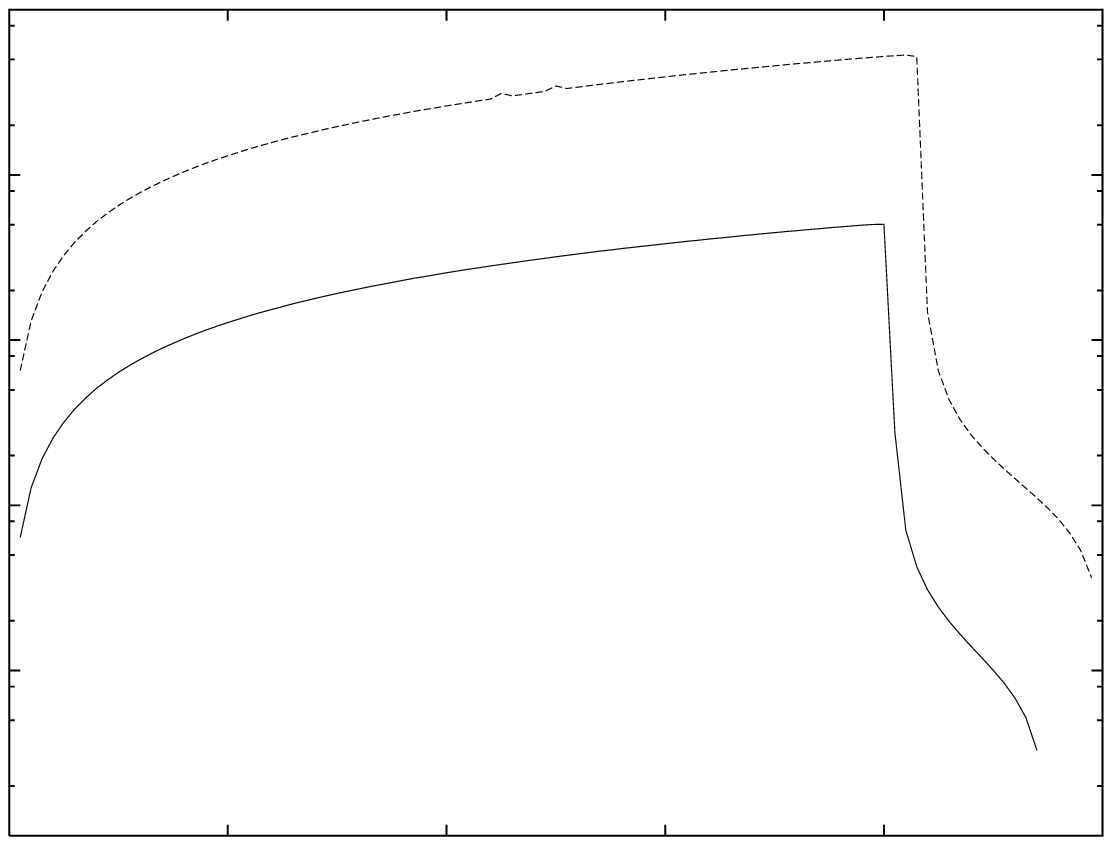,height=7cm,width=12cm}
         }
\put( 0,  1){$10^{-10}$}
\put( 0, 39){$10^{-9}$}
\put( 0, 77){$10^{-8}$}
\put( 0,115){$10^{-7}$}
\put( 0,153){$10^{-6}$}
\put( 0,191){$10^{-5}$}
\put(-60,100){
\LARGE
$\frac{\mbox{d}\Gamma}{~~\mbox{d}m_{\gamma\gamma}}$}
\put(168,-25){\Large $m_{\gamma\gamma}$}
\put( 87,-10){$ 50$}
\put(145,-10){$100$}
\put(203,-10){$150$}
\put(266,-10){$200$}
\put(321,-10){$250$}
\put(350,-10){(MeV)}
\put(60,177){\Large $\phi\rightarrow \rho  \gamma\gamma$}
\put(60,97){\Large $\phi\rightarrow \omega\gamma\gamma$}
\end{picture}
\vskip 1cm
\caption{The $m_{\gamma \gamma}$ spectra in  $\phirho$ (the dashed curve)
and $\phiom$ (the solid curve).}
\label{figtwo}
\end{figure}
\begin{table}
\caption{ $C_{PV}$ defined in Eq.~(7) for $P=\pi^{0}, \eta, \eta^{'}$ and
$V = \rho^{0}, \omega, \phi$.
}
\label{table1}
\begin{tabular}{p{1.3cm}|p{1.9cm}|p{4.4cm}|p{4.4cm}}
\hskip .4cm $C_{PV}$&
\hskip .2cm $\pi$&
\hskip 1.5cm$\eta$&
\hskip 1.5cm$\eta^\prime$\\
\hline
\hskip .5cm $\rho$&
$-\frac{1}{3}$&
\hskip .15cm$ \frac{1}{  \sqrt{3}}(\sqrt{2}\sin\theta-\cos\theta)$&
\hskip .15cm
$-\frac{1}{  \sqrt{3}}(\sqrt{2}\cos\theta+\sin\theta)$\\
\hline
\hskip .5cm $\phi$&
\hskip .3cm$0$&
$ \frac{2}{ 3\sqrt{3}}( \sqrt{2}\cos\theta+\sin\theta)$&
\hskip .3cm $\frac{2}{ 3\sqrt{3}}( \sqrt{2}\sin\theta-\cos\theta)$\\
\hline
\hskip .5cm $\omega$&
$-1$&
$ \frac{1}{ 3\sqrt{3}}( \sqrt{2}\sin\theta-\cos\theta)$&
$-\frac{1}{ 3\sqrt{3}}( \sqrt{2}\cos\theta+\sin\theta)$\\
\end{tabular}
\end{table}
\begin{table}
\caption{ Branching ratios for $V \rightarrow P \gamma$ and $P \rightarrow
V \gamma$ with $P = \pi^{0}, \eta, \eta^{'}$ and $V = \rho^{0}, \omega,
 \phi$.}
\begin{tabular}{p{3cm}|p{4.5cm}|p{4.5cm}}
\hskip .4cm Decay Modes  &   Predictions &
\hskip .4cm Data
\\
\hline
\hskip .7cm
$\omega \rightarrow \pi^{0} \gamma$ & $9.0 \%$  & $(8.5 \pm 0.5) \%$
\\
\hskip .7cm
$\omega \rightarrow \eta \gamma$ & $ 9.5 \times 10^{-4}$
 &  $(8.3 \pm 2.1)\times 10^{-4}$
\\
\hskip .7cm
$\rho^{0} \rightarrow \pi^{0} \gamma $ & $ 5.3 \times 10^{-4}$
 & $(7.9 \pm 2.0) \times 10^{-4}$
\\
\hskip .7cm
$\rho^{0} \rightarrow \eta  \gamma $ & $ 4.1 \times 10^{-4}$
& $ (3.8 \pm 0.7 ) \times 10^{-4}$
\\
\hskip .7cm
$\eta^{'} \rightarrow \rho^{0} \gamma$ & $ 34.3 \% $ & $(30.2 \pm 1.3) \%$
\\
\hskip .7cm
$\eta^{'} \rightarrow \omega \gamma $ & $ 3.1 \% $ & $(3.02 \pm 0.30) \%$
\\
\hskip .7cm
$\phi \rightarrow \pi^{0} \gamma $ & $ 0.0$  & $(1.31 \pm 0.13)
\times 10^{-3}$
\\
\hskip .7cm
$\phi \rightarrow \eta \gamma$ & $ 2.2 \% $ & $(1.28 \pm 0.06) \% $
\\
\hskip .7cm
$\phi \rightarrow \eta^{'}  \gamma$ & $2.1 \times 10^{-4}$ &
$ < 4.1 \times 10^{-4}$
\end{tabular}
\end{table}
\end{document}